# Collapse of Spin-Splitting in the Quantum Hall Effect

M. M. Fogler and B. I. Shklovskii

*Theoretical Physics Institute, University of Minnesota, 116 Church St. Southeast, Minneapolis, Minnesota 55455*

(June 19, 1995 02:56pm)

It is known experimentally that at not very large filling factors $\nu$ the quantum Hall conductivity peaks corresponding to the same Landau level number $N$ and two different spin orientations are well separated. These peaks occur at half-integer filling factors $\nu = 2N + 1/2$ and $\nu = 2N + 3/2$ so that the distance between them $\delta\nu$ is unity. As $\nu$ increases $\delta\nu$ shrinks. Near certain $N = N_c$ two peaks abruptly merge into a single peak at $\nu = 2N + 1$. We argue that this collapse of the spin-splitting at low magnetic fields is attributed to the disorder-induced destruction of the exchange enhancement of the electron $g$-factor. We use the mean-field approach to show that in the limit of zero Zeeman energy $\delta\nu$ experiences a second-order phase transition as a function of the magnetic field. We give explicit expressions for $N_c$ in terms of a sample's parameters. For example, we predict that for high-mobility heterostructures $N_c = 0.9 d n^{5/6} n_i^{-1/3}$, where $d$ is the spacer width, $n$ is the density of the two-dimensional electron gas, and $n_i$ is the two-dimensional density of randomly situated remote donors.

## I. INTRODUCTION

A characteristic feature of the quantum Hall effect is the appearance of peaks in the diagonal conductivity $\sigma_{xx}$ as magnetic field is varied. The conventional argumentation for this is as follows. Consider the density of states diagram for a two-dimensional electron gas (2DEG) in a perpendicular magnetic field (Fig. 1a). It consists of disorder broadened Landau level subbands (LLS): $0\uparrow, 0\downarrow, 1\uparrow, 1\downarrow$, etc. Here arrows stand for the two spin orientations. When the magnetic field is decreased, the Fermi level $\epsilon_F$ consequtively passes through these subbands. Whenever the Fermi energy coinsides with the center of some LLS, where there is a delocalized state, the peak in the dissipative conductivity occurs.

For the case shown at Fig. 1a where the width of LLS is smaller than the distance between them, peaks are positioned at half-integer values of the average filling factor $\nu$: $\nu = 2N + 1 \pm \frac{1}{2}$, where $N = 0, 1, \ldots$ is the Landau level number, and the plus/minus sign corresponds to the spin down/spin up orientation. Experimentally, this periodic positioning of the peaks is observed at small $\nu$ (high fields). In other words, at small $\nu$ two peaks for the same $N$ are well separated. The distance $\delta\nu$ between them is approximately unity. At larger $\nu$ (lower fields) the periodicity of the peak positions is violated. Peaks are found to appear in close pairs at $\nu = 2N + 1 \pm \frac{1}{2}\delta\nu$ with $\delta\nu < 1$. The experiment suggests that starting from a certain $N = N_c$, this separation becomes so small that only single peaks instead of pairs are seen at odd integer $\nu$[1,2].

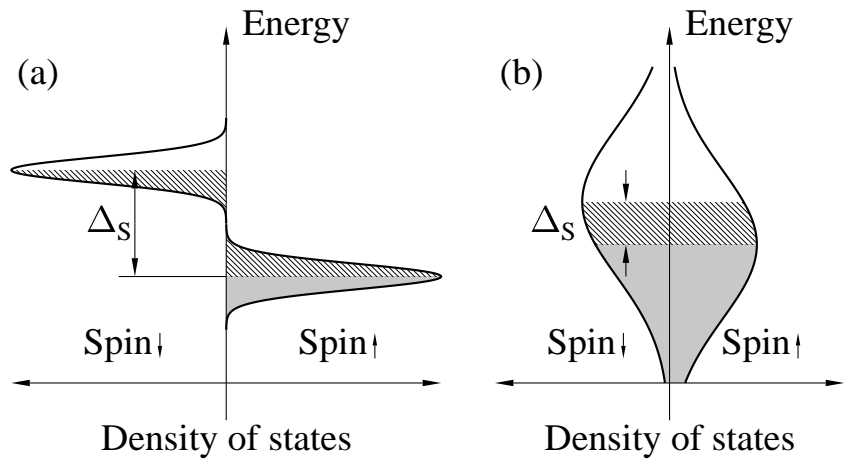

FIG. 1. The density of states diagram where the two upper LLS are shown as disorder-broadened peaks. The ratio of the shaded area to the total area of one LLS peak gives $\delta\nu$. If the width of LLS is smaller then their separation $\Delta_S$ (a), then $\delta\nu$ is close to one. In the opposite case (b) it is much smaller.

Returning back to the density of states diagram, the smallness of $\delta\nu$ can be interpreted as a result of the *overlapping* of LLS (Fig. 1b). The larger the overlap is the smaller the separation between the spin peaks become.

It is worth mentioning that the possibility to observe a *pair* of peaks instead of a single one depends not only on their separation but also on the temperature because as the temperature increases, the conductivity between the peaks rapidly rises. At low temperatures it is, the most likely, due to the variable-range hopping. This mechanism was discussed by Polyakov and Shklovskii[3]. Recently, Polyakov and Raikh[4] studied the conductivity between the peaks in a higher temperature range. In this paper we consider the case of *zero temperature* where peaks are very narrow and concentrate on the peak positions only. However, we have to face the fact that all the experiments are performed at finite temperatures, and small $\delta\nu$ can not be measured accurately. Nevertheless, we assume that the point $\delta\nu = 1/2$ can be reliably identified and define $N_c$ as the smallest peak number where $\delta\nu \geq 1/2$.

It appears from the experiment that $\delta\nu$ is almost constant ($= 1$) away from $N_c$ but decreases rapidly in the very vicinity of $N_c$. Usually, as $N$ changes from, say, $N_c - 1$ to $N_c + 1$, $\delta\nu$ drops from $\delta\nu > 0.75$ to $\delta\nu < 0.25$. For GaAs/Ga$_{1-x}$Al$_x$As heterostructures $N_c$ is usually in the range $1 - 12$, and grows with the mobility and the setback distance of the doped layer[1,2].

In this paper we argue that the sharp change in $\delta\nu$, or in other words, the collapse of the spin-splitting at low magnetic fields, is attributed to the disorder-induced destruction of the exchange enhancement of the electron $g$-factor[5,6]. In GaAs the contribution of the exchange interaction to the energy splitting between the LLS (the exchange gap) exceeds the bare Zeeman energy $Z = g_0 \mu_B B$ by a factor of twenty[7,8] (here $\mu_B$ is the Bohr magneton, $B$ is the magnetic field, and $g_0$ is the bare electronic $g$-factor). This exchange gap can be defined as the energy necessary for creation of a well separated pair of an electron in the spin down and a hole in the spin up LLS. The size of the exchange gap is the largest at odd integer filling factors where in the absence of disorder the spins at the upper LLS are completely polarized. We denote this maximum value by $E_0$ and the ratio $E_0/\hbar\omega_c$ by $\alpha$. The parameter $\alpha$ is related to the $g$-factor via $\alpha = \frac{1}{2}(g - g_0)(m/m_0)$, where $m$ and $m_0$ are the effective and the bare electron masses, respectively. Experimentally[7,8] $g \sim 6\text{-}7$; therefore, $\alpha \sim 0.25$. From the theoretical side, the calculation of the exchange gap for arbitrary Landau level number $N$ was performed already in the early paper of Ando and Uemura[6] (see also the paper by Smith *et al*[9] and references therein). However, the explicit expression was given only recently by Aleiner and Glazman[10] who showed that $\alpha$ becomes independent of $N$ when $N$ gets large:

$$\alpha = \frac{\ln(2k_F a_B)}{\pi k_F a_B}, \quad N \gg k_F a_B \gg 1, \qquad (1)$$

where $k_F$ is the Fermi wave-vector and $a_B$ is the Bohr radius in the semiconductor. In realistic situation $k_F a_B \sim 1$, so Eq. (1) may not apply literally. Nevertheless, for $k_F a_B = 1$ Eq. (1) gives the value of $\alpha \approx 0.22$. In our theoretical arguments we assume that $k_F a_B \gg 1$, but we will keep in mind that this parameter is close to unity when we make our estimates.

So far we have not allowed for a possibility of the spin orientation being different from parallel or antiparallel to the magnetic field. It is known that by limiting our consideration in such a way we can reach incorrect conclusions for $\nu \leq 2$. For example, Sondhi *et al*[11] showed that the lowest energy excitations for $\nu = 1$ are not the electron-hole but skyrmion-antiskyrmion pairs. (The skyrmion is a particular spin texture whose local spin orientation smoothly rotates from a point to point in space, the whole picture resembling a hedgehog. The antiskyrmion is a similar texture but with the opposite direction of rotation). Subsequently, however, Wu and Sondhi[12] found that for larger filling factors ($\nu = 3, 5$) the energy of these excitations $2\Delta_{Sk}$ exceeds that of electron-hole pairs $E_0$. Therefore, already for these filling factors spin textures are not the relevant physical states. Moreover, it can be shown that for large $\nu$ the ratio $2\Delta_{Sk}/E_0$ becomes much larger than unity[10]. For this reason throughout the paper where we are dealing with $N \geq 1$ we assume that the spin orientation can be only up or down.

We argue that in the limit of zero Zeeman energy, in a mean-field approximation $\delta\nu$ experiences a second-order phase transition as a function of the magnetic field. The possibility of such a transition has been mentioned earlier (Ando and Uemura[6], Yarlagadda[13], MacDonald and Yang[14]). However, no explicit predictions for the transition point $N_c$ were made. In order to make such predictions we have to choose realistic models for the disorder potential. Below we consider two limiting cases of short-range and long-range disorder. These cases supposedly describe low-mobility and high-mobility samples, respectively.

In Sec. II we investigate the case of short-range disorder and develop the mean-field description of the phase transition. The number $N_c$ in this case can be expressed in terms of the single-particle scattering time $\tau$:

$$N_c = \frac{2\alpha^2}{\pi\hbar}\epsilon_F \tau \approx \frac{n}{10^{10}\text{cm}^{-2}} \frac{\mu}{10^6 \text{cm}^2/\text{Vs}}, \qquad (2)$$

where in the last equation we used $\alpha = 0.25$ and the fact that for the short-range disorder the single-particle scattering time $\tau$ and the *transport* scattering time $\tau_{tr}$, which enters the expression for the mobility $\mu = e\tau_{tr}/m$, are equal.

Eq. (2) is in agreement with empirical observations that for samples with mobilities in the range $\mu \lesssim$

$50,000 \text{cm}^2/\text{Vs}$ and the 2DEG densities $n$ of order $2.0 \cdot 10^{11} \text{cm}^{-2}$, it is usually only $N = 0, 1$ peaks that are spin-split.

In Sec. III and IV we consider a different model, which applies to high-mobility modulation-doped $\text{GaAs}/\text{Ga}_{1-x}\text{Al}_x\text{As}$ heterostructures. In this model the disorder potential is created by a plane of randomly positioned ionized donors set back from the 2DEG by a distance $d$. The Fourier harmonics of the disorder potential with wavelengths larger than $d$ do not reach the 2DEG. Therefore, for large $d$ the disorder potential is long-range. Its amplitude is determined by fluctuations $\delta n_D(\boldsymbol{r})$ of the two-dimensional donor density around its average value $\langle n_D \rangle$. Due to the Coulomb interactions between charged donors these fluctuations are usually much smaller than in the case of completely random distribution of donors. Following Refs. 15, 16 we assume that the fluctuations can be described by the correlation function

$$\langle \delta n_D(\boldsymbol{r_1}) \delta n_D(\boldsymbol{r_2}) \rangle = n_i \delta(\boldsymbol{r_1} - \boldsymbol{r_2}), \quad (3)$$

where $n_i \ll \langle n_D \rangle$ is the density of "uncorrelated" donors. This new parameter $n_i$ depends on $d$ and a "freeze-out" temperature[15–17], and *is not* explicitly related to the actual donor density.

To find the disorder potential we also have to look at the screening properties of the 2DEG. To this end we has extended the theory of the non-linear screening (see Refs. 15, 18–21) to the case of weak magnetic fields (large $N$). A main new ingredient here was taking into account the screening properties of the lower $N$ completely filled Landau levels.

Using thus calculated disorder potential distribution, we then found the critical number $N_c$. Depending on $n$, it is given by the following three expressions:

$$N_c = \begin{cases} \sqrt{8\pi}\, \alpha^2 n^{3/2} n_i^{-1} = \\ 0.02 d \dfrac{n^{1/2} \ln^2(8\pi n a_B^2)}{n_i a_B^2}, & n \gtrsim n_* \quad (4a) \\[1em] 0.9\, d \dfrac{n^{5/6}}{n_i^{1/3}}, & n_i \lesssim n \lesssim n_* \quad (4b) \\[1em] 0.9\, d \dfrac{n^{2/3}}{n_i^{1/6}}, & n \lesssim n_i, \quad (4c) \end{cases}$$

where $n_*$ is such a value of $n$ that $N_c$ given Eq. (4a) matches the one given by Eq. (4b), i.e., $n_*$ is the solution[22] of the transcendental equation

$$n_*/n_i = [0.4/\alpha(n_*)]^3. \quad (5)$$

Following Ref. 16 we have estimated $n_i$ to be of the order $0.1 \div 1 \cdot 10^{11} \text{cm}^{-2}$, so that $0.01 \lesssim n_i a_B^2 \lesssim 0.1$. This results into the values of $n_* a_B^2$ in the range $0.1 \div 0.3$ or $n_* = 1.0 \div 3.0 \cdot 10^{11} \text{cm}^{-2}$. Having in mind that the experimental range of $n$ is $0.4 \div 4 \cdot 10^{11} \text{cm}^{-2}$, we conclude that Eq. (4b) is probably the most oftenly realized in the experiment. However, for large electron densities Eq. (4a) still may apply.

Our predictions can be verified in detail on gated heterostructures[23]. Changing the gate voltage, it is possible to vary $n$ in a controlled manner and study the different regimes defined by Eqs. (4a-4c). Changing both the gate voltage and the magnetic field, one can fix $N$ and study the collapse of a particular spin-split peak[24]. This may be a better way to study the critical behavior experimentally compared to the examining of $\delta\nu$ at discreet values $N$.

Concluding the introduction, we note that the collapse of the spin-splitting can be represented in terms of the global phase diagram for the quantum Hall effect originally suggested by Kivelson *et al*[25] for the spinless case. The modified diagram is shown in Fig. 2. The purpose of this paper is to find the equation for the dashed line in this figure.

The paper is organized as follows. In Sec. II we present the mean-field treatment of the problem and apply the results to the case of low-mobility samples. In Sec. III we study the case of high-mobility heterostructures with large electron densities: $n/n_i \gg \alpha^{-3}$. We show that in this case the mean-field description is adequate. In Sec. IV we complement the mean-field picture by a microscopic consideration and then derive from that the expression for $N_c$ in the case of moderate and small electron densities $n/n_i \ll \alpha^{-3}$. Finally, in Sec. V we make several concluding remarks.

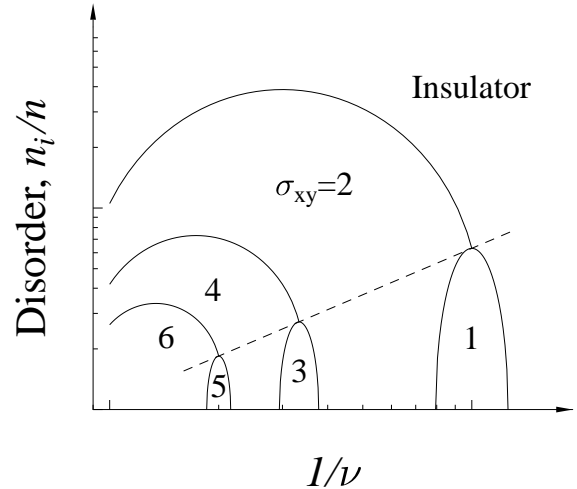

FIG. 2. The global phase diagram for the quantum Hall effect as obtained in the mean-field approximation (schematically). The Zeeman energy is neglected. The dashed line is defined by the relation $\nu = 2N_c + 1$ with $N_c$ given by Eqs. (4a-4c).

## II. MEAN-FIELD DESCRIPTION OF THE PHASE TRANSITION FOR THE SPIN-SPLITTING

Our goal is to study the phase transition in the spin degree of freedom in the system of interacting electrons in weak magnetic field and in the presence of an external disorder potential. As we will see this phase transition occurs when the amplitude of the disorder is still less than the cyclotron gap $\hbar\omega_c$ and therefore, only the two upper LLS are partially filled while all the lower $2N$ LLS are completely occupied. In this situation all the phenomena of interest take place at the upper Landau level. Therefore, a natural idea is to describe the electrons at the upper LLS by means of some effective Hamiltonian where the degrees of freedom of the lower LLS are integrated out. This program was realized in Ref. 10. It uses the idea that the bare Coulomb interaction of the electrons at the upper LLS is strongly reduced due to the screening by lower LLS[26]. The fact that interaction is small suggests using perturbative methods, e.g, the Hartee-Fock approximation.

Consider an electronic state at the upper LLS in the form of a Hartree-Fock function based on one-electron wave-functions proposed by Kivelson et al[27] (so-called coherent states). We will not need the explicit form of these wave-functions. It is sufficient to know that for large $N$ the square of the absolute value (the probability density) of each wave-function of this type is not small only within a very narrow ring of width $k_F^{-1}$ and radius $R = \nu/k_F$ (the classical cyclotron radius). For a "ring" with the center at point $r_0$ the probability density at a point $r$ is $F(r - r_0)$, where

$$F(r) \approx \frac{1}{2\pi R}\delta(r - R). \quad (6)$$

is the form-factor of the wave-functions. The ring centers must be chosen such that their density be equal to $n_L = 1/(2\pi l_B^2)$, where $l_B = \sqrt{\nu}/k_F$ is the magnetic length. Define the local filling factors $\nu_\uparrow(r)$ and $\nu_\downarrow(r)$ as the fractions of occupied states with centers near the point $r$, and also $\nu_N = \nu_\uparrow(r) + \nu_\downarrow(r)$ as the total filling of the upper Landau level.

Our effective Hamiltonian has the form

$$H = H_{\text{int}} + H_{\text{imp}} + H_Z, \quad (7)$$
$$H_{\text{int}} = H_{\text{ex}} + H_H, \quad (8)$$

where all the terms are expressed via the local filling factors. For example, the exchange energy of the system $H_{\text{ex}}$ is

$$H_{\text{ex}} = -\frac{1}{2}n_L^2 \int d^2r d^2r' G_{\text{ex}}(r - r') \times [\nu_\uparrow(r)\nu_\uparrow(r') + \nu_\downarrow(r)\nu_\downarrow(r')]. \quad (9)$$

The kernel of this expression, $G_{\text{ex}}(r)$, is the energy of the exchange interaction of two "rings" whose centers are a distance $r$ apart. The Hartree energy of the system $H_H$ is given by

$$H_H = \frac{1}{2}n_L^2 \int d^2r d^2r' \nu_N(r) G_H(r - r')\nu_N(r'), \quad (10)$$

where $G_H(r)$ is now the energy of the direct interaction of two "rings". The term $H_{\text{imp}}$ describes the interaction with random impurities:

$$H_{\text{imp}} = n_L \int d^2r d^2r' U(r') F(r' - r)\nu_N(r), \quad (11)$$

where by $U(r)$ we mean the impurity potential already screened by the lower completely filled LLS. The form-factor $F$ here [see Eq. (6)] accounts for the fact that the energy of each one-particle state is determined by the potential *averaged* over the the cyclotron orbit[28]. To simplify the formulae we introduce an additional notation for this averaged potential:

$$V(r) = \int d^2r' U(r') F(r' - r), \quad (12)$$

so that

$$H_{\text{imp}} = n_L \int d^2r V(r)\nu_N(r), \quad (13)$$

Finally, there is the Zeeman term

$$H_Z = \frac{1}{2}n_L Z \int d^2r [\nu_\downarrow(r) - \nu_\uparrow(r)]. \quad (14)$$

To complete the definition of the Hamiltonian we also have to provide the explicit expressions for $G_{\text{ex}}(r)$, $G_H(r)$ and $U(r)$.

According to Ref. 10

$$G_{\text{ex}}(r) = \frac{e^2 a_B}{2\pi R}\frac{1}{r(1 + r/\xi)}, \quad l_B \ll r < 2R, \quad (15)$$

where

$$\xi = \frac{2}{k_F a_B}R. \quad (16)$$

$G_{\text{ex}}(r)$ rapidly falls off beyond the distance $\xi$, which can be called the range of the exchange interaction. It is easy to verify that if $N \gg (k_F a_B)^2$, then $\xi \gg l_B$. Hence, for such $N$ the number of electrons at the upper Landau level involved in the exchange interaction with the given one is large and we can indeed describe the distribution of electrons by continuous functions $\nu_\uparrow(r)$ and $\nu_\downarrow(r)$. In the experimental situation $k_F a_B \sim 1$, and therefore, we are referring to $N \gg 1$.

The explicit formulae for $G_H(r)$ and $U(r)$ will not be used in the present Section. We postpone their discussion until Sec. III.

The ground state distributions of $\nu_\uparrow(r)$ and $\nu_\downarrow(r)$ can be inferred from the fact that $H$ has the lowest value for a given total number of particles, which gives

$$\nu_\uparrow(r) = f[W(r) - Z/2 - E_\uparrow(r)] \quad (17)$$
$$\nu_\downarrow(r) = f[W(r) + Z/2 - E_\downarrow(r)] \quad (18)$$

where

$$W(\mathbf{r}) = V(\mathbf{r}) + n_{\mathrm{L}} \int \mathrm{d}^2 r' G_{\mathrm{H}}(\mathbf{r}-\mathbf{r}')\nu_N(\mathbf{r}') \qquad (19)$$

is the potential, which includes the screening effects of both the lower completely filled LLS (the first term) and the upper partially filled ones (the second term), $f(\epsilon)$ is the Fermi-Dirac distribution function, and $E_\uparrow$ is given by

$$E_\uparrow = \int \mathrm{d}^2 r' G_{\mathrm{ex}}(\mathbf{r}-\mathbf{r}')\nu_\uparrow(\mathbf{r}') \qquad (20)$$

(and similarly for $E_\downarrow$). Notice that for every position $\mathbf{r}$ of the ring center there are two distinct values of energy $W(\mathbf{r}) - Z/2 - E_\uparrow(\mathbf{r})$ and $W(\mathbf{r}) + Z/2 - E_\downarrow(\mathbf{r})$ corresponding to the two orientations of the electronic spin. They define Landau level subbands. The energy separation $\Delta_{\mathrm{S}}(\mathbf{r})$ between the subbands is a sum of two terms:

$$\Delta_{\mathrm{S}} = Z + E, \qquad (21)$$
$$E = E_\uparrow - E_\downarrow. \qquad (22)$$

Eqs. (17,18) are in agreement with the density of states diagram (Fig. 1), which we discussed above. Using Eqs. (20,22) we can rewrite $E$ in the form

$$E(\mathbf{r}) = \int \mathrm{d}^2 r' G_{\mathrm{ex}}(\mathbf{r}-\mathbf{r}')[\nu_\uparrow(\mathbf{r}') - \nu_\downarrow(\mathbf{r}')]. \qquad (23)$$

If both $\nu_\uparrow(\mathbf{r})$ and $\nu_\downarrow(\mathbf{r})$ are uniform on the scale of $\xi$, then $E$ is proportional to the difference in the subband populations[6]:

$$E = E_0(\nu_\uparrow - \nu_\downarrow), \qquad (24)$$

where $E_0 = \int \mathrm{d}^2 r G_{\mathrm{ex}}(\mathbf{r}) = \alpha \hbar \omega_c$. Performing this integration with the help of Eq. (15), we recover Eq. (1)[29].

Since we are interested in the conductivity peaks, from now on by $\nu_\uparrow$ and $\nu_\downarrow$ in Eq. (24) we will understand the corresponding quantities for the position of the Fermi level at the center of $N \uparrow$ LLS as shown in Fig. 1. Eq. (24) is one of the central equations in the mean-field theory of the spin-splitting phase transition. Another equation relates $\nu_\uparrow - \nu_\downarrow$ to the separation $\delta\nu$ between the conductivity peaks:

$$\nu_\uparrow - \nu_\downarrow = \frac{1}{2}\delta\nu. \qquad (25)$$

Indeed, it is easy to see that $\delta\nu$ is the total area of the shaded strip (of width $\Delta_{\mathrm{S}}$) at Fig. 1, whereas $\nu_\uparrow - \nu_\downarrow$ is only a half of this area.

One more equation relates $\delta\nu$ to $\Delta_{\mathrm{S}}$:

$$\delta\nu = \int_{-\Delta_{\mathrm{S}}}^{\Delta_{\mathrm{S}}} \rho(\epsilon)\mathrm{d}\epsilon, \qquad (26)$$

thus completing the system of the mean-field equations Eqs. (21,24,25,26). Here $\rho(\epsilon)$ is a new quantity, which we call the form-factor of LLS. It has the dimensionality of inverse energy and has the direct relation to the density of states in the potential $W(\mathbf{r})$. Namely, the contribution to the to total density of states from, say, $N \uparrow$ LLS centered at the energy $-E_\uparrow$ is given by $n_{\mathrm{L}}\rho(\epsilon + E_\uparrow)$.

Using Eqs. (21,24,25) we arrive at

$$\Delta_{\mathrm{S}} = \frac{1}{2}E_0\delta\nu + Z. \qquad (27)$$

The first (exchange) term is much larger than the second (Zeeman) one, provided $\delta\nu$ is not too small. First, let us neglect the Zeeman energy. In this approximation the collapse of the spin-splitting can be seen the most vividly. We examine the effect of a finite Zeeman energy a bit later.

We are going to solve the system of two equations (26) and (27). This procedure can be illustrated graphically (Fig. 3). We have a non-zero solution for $\delta\nu$ only if $E_0 > \rho^{-1}(0)$. Therefore, the transition point is defined by

$$E_0 = \rho^{-1}(0). \qquad (28)$$

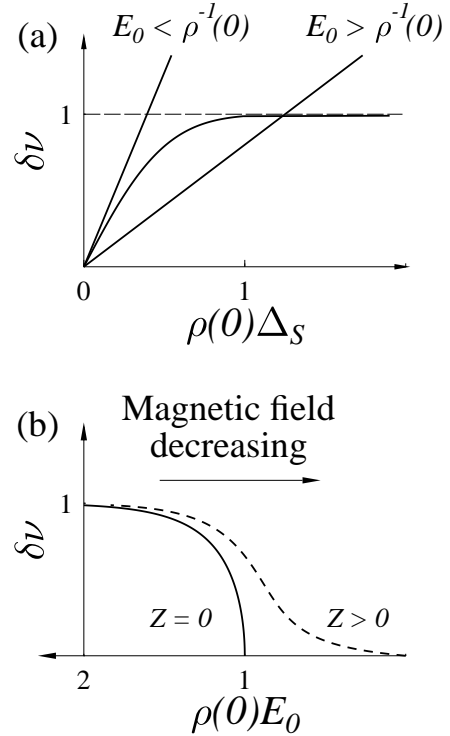

FIG. 3. (a) $\delta\nu$ is given by the intersection of the bold lines representing Eqs. (26) and (27). Non-zero solution exists only for $E_0 > \rho^{-1}(0)$. (b) The dependence of $\delta\nu$ on $E_0$ has a typical second-order phase transition form. Non-zero $Z$ smears the transition. To generate these plots we used the Gaussian form for the density of states: $\rho(\epsilon) = \rho(0)\exp(-\pi\rho^2(0)\epsilon^2)$. The solid and the dashed lines correspond to $Z = 0$ and $Z = 0.06 E_0$, respectively.

Near the transition $\delta\nu$ can be estimated using Taylor expansion for $\rho(\epsilon)$ in Eq. (26):

$$\delta\nu = \left[\frac{-24}{\rho''(0)E_0^3}\right]^{1/2} \left[\frac{E_0 - \rho^{-1}(0)}{E_0}\right]^{1/2}. \quad (29)$$

It is easy to see that our approach is very similar in spirit to such well-known mean-field theories as Stoner model of band ferromagnetism[30] or Weiss' theory of the molecular field. In the latter case the analogy is achieved via substitutions "disorder" ($\rho^{-1}$) $\rightarrow$ temperature[31], exchange energy ($E$) $\rightarrow$ magnetization. Following this analogy, the Zeeman term is mapped onto the external field. It is clear now that the effect of this term is the smearing of the transition (Fig. 3b). For instance, in the "paramagnetic phase" we have an analogue of Curie-Weiss law:

$$\delta\nu = \frac{2Z}{\rho^{-1}(0) - E_0}, \quad \rho^{-1}(0) - E_0 \gg Z. \quad (30)$$

Now let us address the question how reliable the mean-field approach is. It is based on the assumption that $\nu_\uparrow$ and $\nu_\downarrow$ can be treated as spatially uniform. In reality, they fluctuate from point to point and in the immediate vicinity of the transition where these fluctuations exceed the average value of $\nu_\uparrow - \nu_\downarrow$, the mean-field theory definitely breaks down. However, in this paper we do not wish to discuss the details of the critical behavior at small $\delta\nu$. Instead, we focus on not too small $\delta\nu$, say, $\delta\nu \geq 1/2$. The characteristic scale of the fluctuations in $\nu_\uparrow$ and $\nu_\downarrow$ is set by the correlation length $l_{cor}$ of the disorder potential. As long as $l_{cor} \ll \xi$, the fluctuations in $\nu_\uparrow - \nu_\downarrow$ are averaged out, and therefore, the mean-field theory is a very good approximation. For example, it is expected to work for low-mobility samples where the disorder is short-range, i.e., $l_{cor} \ll l_B$ (and we assume $N \gg (k_F a_B)^2$ so that $l_B \ll \xi$ as explained above). In this case we can express $N_c$ in terms of the sample mobility. Indeed, it has been shown[28] that for the short-range disorder the SCBA approximation is valid. In this approximation

$$\rho(0) = \hbar^{-1}\sqrt{(2/\pi)(\tau/\omega_c)}, \quad (31)$$

Together with Eq. (28) this leads to Eq. (2), which as we mentioned in the Sec. I, qualitatively agrees with the experimental data.

In the next Section we discuss the case of high-mobility heterostructures with $n/n_i \gg \alpha^{-3}$. As we will see, the simple theory presented in this Section gives the adequate description of the phase transition.

### III. HIGH-MOBILITY HETEROSTRUCTURES: LARGE ELECTRON DENSITIES

In the preceding Section we applied our mean-field approach to low-mobility samples where the disorder potential is short-range. Let us now turn to the case of high-mobility samples. In particular, we consider a heterostructure described by a set of three parameters $n$, $n_i$, and $d$ as discussed in Sec. I. In this Section we will find the transition point $N_c$ assuming $n/n_i \gg \alpha^{-3} \gg 1$.

At first sight the problem is different from the one in the preceding Section because the disorder potential created by remote donors is now long-range with the correlation length of order $d \gg l_B$. This length also sets the scale for the variations of $\nu_\uparrow$ and $\nu_\downarrow$ [Eqs. (17,18)]. However, we will show below that the transition point is given by Eq. (4a); therefore, near the transition the range of the exchange interaction $\xi \sim 2\pi\alpha R \sim \alpha^3 dn/n_i$ is much larger than $d$, and so on the scale of $\xi$ both $\nu_\uparrow$ and $\nu_\downarrow$ are uniform and our mean-field treatment still applies. The problem now reduces to the calculation of the r.h.s of Eq. (28), i.e., to the calculation of the LLS width. This width is nothing else than the amplitude of $W(\boldsymbol{r})$. More precisely,

$$\rho(0)^{-1} = \sqrt{2\pi\langle W^2 \rangle} \quad (32)$$

if $W(\boldsymbol{r})$ has the normal distribution[28]. To calculate this quantity we need to know what the potential created by fluctuating donor density is, and how it is screened by the 2DEG. The former is characterized by exponentially suppressed short-scale harmonics (with the wave-vectors larger than $d^{-1}$). The screening in its turn greatly reduces the amplitudes of large-scale harmonics. As a result, the harmonics with $q \sim d^{-1}$ dominate in the potential, and so we can focus only on the screening at these $q$.

Both the lower, completely filled LLS (the incompressible or i- liquid) and the upper, partially filled LLS (the compressible or c- liquid) participate in the screening, which is reflected in the fact that $W(\boldsymbol{r})$ as given by Eq. (19) is a sum of two terms. The screening by the i-liquid is described by the dielectric constant[10,26]

$$\varepsilon = \varepsilon_s \times \begin{cases} 1 + R^2 q/a_B, & q \ll 1/R & (33a) \\ 1 + 2/a_B q, & 1/R \ll q \ll k_F, & (33b) \end{cases}$$

where $\varepsilon_s$ is the dielectric constant of the semiconductor. In the asymptotic expression for large $q$ [Eq. (33b)] one recognizes the familiar zero magnetic field result[32]. This is natural because large $q$ correspond to small distances $q^{-1} \ll R$ at which the curvature of the cyclotron orbits may not be important. The dominant harmonics with $q \sim d^{-1}$ belong to this range because near the transition $\xi \gg d$ and, of course, $R \gg \xi$, so that $R \gg d$. We will show that the contribution to the screening of the dominant harmonics from the c-liquid is negligible, and so

$$W(q) \approx V(q), \quad q \sim d^{-1}. \quad (34)$$

Therefore, for calculation of $\sqrt{\langle W^2 \rangle}$ we will need to consider only the screening by the lower LLS and only in the regime $q \gg 1/R$ described by Eq. (33b). Remind that $V$ is the potential averaged over the cyclotron orbits and in real space it is expressed in terms of the unaveraged potential $U$ by means of Eq. (12). Now we rewrite this equation in $q$-space:

$$V(q) = F(q)U(q). \qquad (35)$$

where $F(q)$ can be deduced from Eq. (6) to be

$$F(q) = J_0(qR), \quad q \ll k_{\rm F}. \qquad (36)$$

Due to the averaging over the cyclotron orbits, the amplitude of $V(\mathbf{r})$ is reduced compared to that of $U(\mathbf{r})$ by a factor of order $\sqrt{d/R}$[28]. This is because every cyclotron orbit samples approximately $R/d$ uncorrelated areas of size $d$ each. Another way to see that is to notice that for the dominant harmonics $q \sim d^{-1}$ we have $|F(q)| \sim \sqrt{d/R}$.

From Eqs. (32,35) we obtain

$$\rho(0)^{-2} = \int_0^\infty {\rm d}q\, q J_0^2(qR) S(q), \qquad (37)$$

where $S(q)$ is the structure factor of the disorder potential defined via

$$\langle U(\mathbf{q_1})U(\mathbf{q_2})\rangle = (2\pi)^2 S(q)\delta(\mathbf{q_1}+\mathbf{q_2}). \qquad (38)$$

Using this definition and Eq. (3) we find

$$S(q) = \left[\frac{2\pi e^2 n_i^{1/2}}{q\varepsilon(q)}{\rm e}^{-qd}\right]^2. \qquad (39)$$

so that

$$\frac{1}{\rho(0)} = \sqrt{2\pi\langle W^2\rangle} = \sqrt{\frac{\pi}{2}}\,\frac{e^2 a_{\rm B} n_i^{1/2}}{d}\left(\frac{d}{R}\right)^{1/2}. \qquad (40)$$

Together with Eq. (28) this leads to

$$N_c = 2\alpha^2 k_{\rm F} dn/n_i, \qquad (41)$$

which is the same as Eq. (4a).

Now we would like to justify our assumption that the screening by the c-liquid is negligible. We will show that this is true provided $n/n_i \gg \alpha^{-3}$.

The screening by the c-liquid may be both in the linear and the non-linear regimes depending on the amount of this liquid. This amount varies with the filling factor $\nu$, which gives rise to numerous screening regimes in different regions of the global phase diagram of Fig. 2. Since we are interested primarily in the positions of the conductivity peaks, we will focus here on a small part of this diagram near the phase boundaries defined by the equation $\nu = 2N_c + 1 \pm \frac{1}{2}\delta\nu(n_i,n)$. As we move along such a boundary in the direction of the increasing disorder, we proceed from the point where the amplitude of $W$ is zero and the spin-splitting is the largest, to the point where this amplitude is $\alpha\hbar\omega_c$ and the spin-splitting vanishes. Consequently, all the way along this path the amplitude of $W$ never exceeds $\alpha\hbar\omega_c$ and moreover $\hbar\omega_c$. This means that as long as $N < N_c$, the c-liquid forms only at the two upper LLS (for the exception of rare places where $W(\mathbf{r})$ is untypically large).

To estimate the (averaged) potential created by the c-liquid we need to know the expression for $G_{\rm H}(\mathbf{r})$. It is given by

$$G_{\rm H}(\mathbf{r}) = \int\frac{{\bf d}^2{\bf q}}{(2\pi)^2}\frac{2\pi e^2}{q\varepsilon(q)}F^2(q){\rm e}^{{\rm i}\mathbf{q}\cdot\mathbf{r}}, \qquad (42)$$

where we took into account the form-factor of the wave-functions [Eq. (36)] and the screening by the $2N$ completely filled lower LLS [Eq. (33)]. This integral was evaluated in Ref. 10 to be

$$G_{\rm H}(r) = \frac{e^2 a_{\rm B}}{2\pi Rr} + \frac{3e^2 a_{\rm B}}{4\pi^2 R^2}\ln\left(\frac{R}{r}\right) + \frac{e^2 a_{\rm B}}{R^2}\ln\left(\frac{R}{a_{\rm B}}\right),$$
$$l_{\rm B} \ll r \ll R. \qquad (43)$$

Among these three terms, the last one is just a constant and will have no effect in our consideration, and the second term is smaller than the first one. Retaining just the first term we discover a remarkable fact: the renormalized interaction is equivalent to the Coulomb law with an effective dielectric constant

$$\varepsilon_* = \frac{2\pi R}{a_{\rm B}}. \qquad (44)$$

Now the amplitude of the potential created by the c-liquid can be easily estimated. The amplitude of the variations in $\nu_N(\mathbf{r})$ does not exceed one; therefore, the amplitude of the potential is of the order of $e^2 n_{\rm L} d/\varepsilon_* \sim e^2 a_{\rm B} n^{3/2} d/N^2$, which is much smaller than $\sqrt{\langle W^2\rangle}$ if $n/n_i \gg \alpha^{-3}$ and $N \sim N_c$ [see Eqs.(40,41]. Therefore, the screening is performed mainly by the i-liquid as we claimed above[33].

The expressions for $N_c$ obtained for the short-range disorder potential considered in Sec. II and in the present case appear to be different. Let us show that, in fact, they can be cast in a very similar form. Indeed, Eq. (37) can be rewritten in the form

$$\rho(0)^{-1} = \hbar\sqrt{\omega_c/\tau}, \qquad (45)$$

which differs from the short-range case [Eq. (32)] only by a numerical factor $\sqrt{2/\pi}$. In this expression $\tau$ is again the single-particle scattering time at $B = 0$:

$$\frac{\hbar}{\tau} = \frac{m}{\hbar^2 k_{\rm F}}\int_0^\infty {\rm d}q\, S(q). \qquad (46)$$

With the help of Eqs. (28,45) $N_c$ can now be presented in the form almost identical to Eq. (2):

$$N_c = \frac{\alpha^2}{\hbar}\epsilon_{\rm F}\tau. \qquad (47)$$

Therefore, the difference between the two considered cases is largely in the choice of measurable sample parameters, through which we express $N_c$. While for the short-range disorder potential we can extract $\tau$ from the mobility, in the long-range disorder potential case we use $n$, $n_i$, and $d$.

Concluding this Section, we would like to mention that $\tau$ can be estimated by analyzing the Shubnikov-de Haas (SdH) oscillations. Existing theories[32,34,35] predict that the deviation of the resistivity in a weak magnetic field from its zero field value should behave as

$$\frac{\rho_{xx} - \rho_{xx}^0}{\rho_{xx}^0} = -A e^{-\pi/\omega_c \tau} \frac{\frac{2\pi^2 T}{\hbar \omega_c}}{\sinh\left(\frac{2\pi^2 T}{\hbar \omega_c}\right)} \cos\left(\frac{2\pi \epsilon_F}{\hbar \omega_c}\right). \quad (48)$$

There is a disagreement in literature about the value of the numerical coefficient $A$ (see the discussion in Ref. 35) and problems with fitting the experimental data to the theoretical predictions[36]. Nevertheless, it is probably reasonable to assume that the peak number $N_{\rm SdH}$ where the extrapolated to $T = 0$ value of the r.h.s. of Eq. (48) becomes of order unity corresponds to $\omega_c \tau \sim \pi$, i.e., $N_{\rm SdH} \sim \epsilon_F \tau / \pi \hbar$. Comparing with Eq. (47) we obtain

$$N_c \sim \alpha^2 \pi N_{\rm SdH}. \quad (49)$$

With $\alpha = 0.25$ this gives $N_c \sim N_{\rm SdH}/5$ in reasonable agreement with the data by Coleridge et al.[2] where $N_c = 12$ and $N_{\rm SdH} \sim 40$. The reason why there is such a simple relation between $N_c$ and $N_{\rm SdH}$ is that the underlying physics is similar: the SdH oscillations develop when the width of the LLS becomes comparable with the energy separation $\hbar \omega_c$ of the Landau levels, and the collapse of spin-splitting occurs when this width is of the order of the spin subband separation $\alpha \hbar \omega_c$, which differs only by a numerical factor.

In the following Section we are going to complement the mean-field description we employed so far by the consideration on the microscopic level. This will enable us to find $N_c$ for the case of moderate (and small) electronic densities $n \ll n_i/\alpha^3$.

## IV. HIGH-MOBILITY HETEROSTRUCTURES: MODERATE AND SMALL DENSITIES

Consider now the case of moderate electron densities $1 \ll n/n_i \ll \alpha^{-3}$, where the predicted transition point is given by Eq. (4b). One can verify that near the transition we still have the inequality $d \ll R$ so that the averaging over the cyclotron orbits is still important for the calculation of the macroscopic density of states[28]. However, the range of the exchange interaction $\xi \sim \alpha R$, is shorter than the lengthscale on which this density of states is formed ($d$). In this case the mean-field theory we used so far becomes insufficient, and there is a need for a new, local theory.

A first step towards constructing such a theory will be to deepen our understanding of the already studied large density case. The starting point of our analysis was to associate the peaks in $\sigma_{xx}$ with the presence of the delocalized ("metallic") states near the Fermi energy. Now we have to somehow relate these metallic properties to local values of the filling factor $\nu(\mathbf{r})$, which can be either integer (i-liquid) or non-integer (c-liquid). Clearly, the i-liquid where there is no gapless excitations, is unable to carry a dissipative current. It is "insulating" in this sense. Note that in the large density case the i-liquid occupies almost the entire area of the sample. This is due to the fact that near the transition only the two upper LLS are typically partially filled and the potential created by the electrons at these LLS is small. The c-liquid exists only in the form of narrow channels between neighboring "islands" of the i-liquid. These channels are referred to as "bulk edge channels". The peaks in $\sigma_{xx}$ appear when at certain values of the average $\nu$, the bulk edge channels form a percolating network, which can conduct the dissipative current. The transport properties of such a network have been considered by several authors[37,38]. For us it is important to identify the difference in the structure of this network for the "ferromagnetic" and "paramagnetic" phases. This difference is illustrated in Fig. 4a where we show the structure of the channel network for the periodic external potential

$$U(x, y) = U_0 \sin\left(\frac{x + y}{2d}\right) \sin\left(\frac{x - y}{2d}\right), \quad (50)$$

corresponding to the averaged potential $V(\mathbf{r})$ of the same form:

$$V(\mathbf{r}) = U(\mathbf{r}) J_0(R/d). \quad (51)$$

It follows from Eqs. (17,18) that in the "ferromagnetic" phase where $\Delta_S > 0$, the total area occupied by the i-liquid is split into *three* regions where $\nu_N(\mathbf{r})$ takes values 0 (both spin up and spin down LLS are empty), 1 (the spin up LLS is occupied and spin down LLS is empty), and 2 (both LLS are occupied). The bulk edge channel network consists of two subnetworks with different spin orientations: one is the channels with $0 < \nu_N < 1$ (spin up) and the other is those with $1 < \nu_N < 2$ (spin down). The two subnetworks are disconnected as they are the two boundaries of region "1". The percolation through the two networks is achieved at different average $\nu$, and this is why there are two peaks in $\sigma_{xx}$ for a given $N$. Let us find when these peaks occur. Note that the spin up and spin down channels follow the contours of constant $V(\mathbf{r})$ (or, more precisely, of constant $W(\mathbf{r})$ but in this case they are the same). Only one such contour, namely, $V = 0$ percolates through the sample ($V(\mathbf{r})$ is symmetrically distributed around zero). Therefore, the peaks in $\sigma_{xx}$ correspond to $\epsilon_F = -E_\uparrow$ and $\epsilon_F = -E_\downarrow$, which are the centers of the spin up and spin down LLS [see Eqs. (17,18)]. Thus, the network picture is consistent with the one we used previously, and gives an important insight into the nature of the delocalized states at the centers of the LLS.

On the contrary, in the "paramagnetic" phase where $\Delta_S = 0$ (Zeeman energy neglected) only the regions "0" and "2" are present (Fig. 4b). In this case the spin up and spin down networks coincide spatially, and the percolation through the two is achieved simultaneously, which means that $\delta\nu$ is zero.

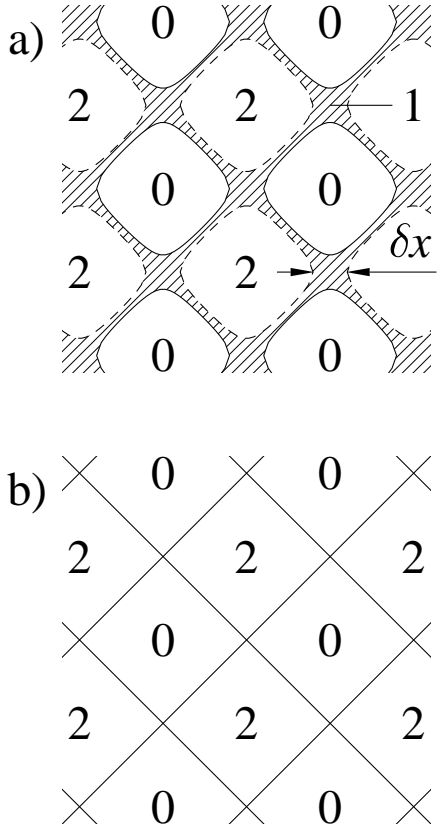

FIG. 4. The distribution of $\nu(\boldsymbol{r})$. Labels "0", "1", and "2" correspond to filling factors $2N$, $2N+1$, and $2N+2$, respectively. (a) "Ferromagnetic phase" ($\delta\nu > 0$). The spin up bulk edge states (solid lines) are disconnected from the spin down ones (dashed lines). The space between them is region "1" (shaded). (b) "Paramagnetic phase" ($\delta\nu = 0$). Only the regions "0" and "2" are present and the two networks spatially coincide.

In Sec. II and III we found that the transition is driven by the competition between the exchange interaction and the disorder. Now we can trace this competition on the microscopic level. The disorder is indiscriminative towards the electron spin. It tries to push the electrons, spin up and spin down all the same, away from the hills of the potential and pile them up in the valleys, thus tending to create doubly-empty ("0") and doubly-filled ("2") regions only. The exchange interaction, on the other hand, prefers the ferromagnetic spin ordering and leads to spin-flip processes at the spin down LLS. Since in the region "2" all the spin up states are occupied, these spin-flip processes occur via filling of the region "0" at the cost of an increase in energy due to the disorder potential. Thus, the region "1" is created, and the spin up and spin down bulk edge channels become spatially separated.

In our example of the chess-board potential the percolating constant energy contour is the one, which goes through the saddle points. Therefore, the properties of the channel networks near the saddle points are extremely important for the percolation (this is true for the random external potential as well). In fact, it is near a saddle point that the spin up and the spin down channels split: one turns in the clockwise and the other in the counterclockwise direction (Fig. 4a). One can say that on the microscopic level the spin-splitting transition is the transition of the saddle points from a zero magnetization "0"-"2" state to a magnetized "0"-"1"-"2" state. Recall that $\delta\nu$ differs from the average magnetization only by a factor of two [Eq. (25)]; therefore, $\delta\nu$ is approximately given by the the ratio of the magnetized area at a single saddle point to the area of a chess-board cell. The former can be estimated as the square of the distance $\delta x$ between the two regions "2" (or two regions "0") at the saddle point in the configuration shown in Fig. 4 (corresponding to $\nu = 2N + 1$). This leads to

$$\delta\nu \sim (\delta x/d)^2, \qquad (52)$$

We see that for $\delta\nu = 1/2$ one should have $\delta x$ of order $d$. We will use this fact later in this Section.

Before we start to analyze the moderate electron density case let us give another explanation why the mean-field description is expected to work in the large $n$ case. The point is that in the "ferromagnetic" phase each magnetized saddle point creates the "exchange field", which supports the existence of its own magnetization and the magnetization of all neighboring saddle points. In the case of large $n$ where the distance between the saddle points ($d$) is much smaller than the range of the interaction ($\xi$), each magnetized saddle point is supported by the field of large number of other saddle points. Therefore, this is a collective phenomenon and the mean-field approach is a good one.

Let us now resume the study of the moderate electron density case. In this case $\xi \ll d$ and each saddle point has to support its magnetization individually. Another difference from the large density case is that the potential created by the electrons at the upper LLS is not small, and the screening is now accomplished mainly by the upper LLS. In other words, both the Hartree and the exchange terms of the electron-electron interaction at the upper LLS are working against the disorder. The important rôle of the Hartree term was emphasized by Dempsey et al[39] who studied a similar "0"-"2" to "0"-"1"-"2" transition for $N = 0$ in an edge of a wide quantum wire. (See also the paper by Manolescu and Gerhardts[40] and references therein). The corresponding effect in our geometry is the "0"-"2" to "0"-"1"-"2" transition on the sides of the chess-board cells. We remind that our interest is primarily in the saddle points. However, our treatment of the problem is similar in spirit; therefore, we briefly review the results obtained in Ref. 39. Let $x$ be the coordinate in the direction transverse to the wire and the edges of the wire be located at $x = 0$ and $x = -b$. We denote the potential providing the lateral confinement of the wire by $V(x)$ (this choice reflects the similarity between this potential and the potential $V(x, y)$ in our chess-board model). The evolution of the edge structure as $V(x)$ is softened is as follows. For large confining fields $dV/dx$ the edge is in "0"-"2" state ($\nu = 2$ in the interval $-b < x < 0$ and $\nu = 0$ everywhere else, see Fig. 5a). When the confining field gets smaller than the Hartree field given by

$$2n_{\rm L} \int_{-b}^{0} dX \int dY \left. \frac{\partial G_{\rm H}(x - X, Y)}{\partial x} \right|_{x=0}, \qquad (53)$$

the region "1" arises spontaneously[39] and then grows (Fig. 5b). For a while the edge density profile remains step-like, i.e., the edge channels are infinitesimally narrow. However, when the width of the region "1" becomes of order $l_{\rm B}$, the edge undergoes a second transition: the edge channels acquire a finite width (Fig. 5c). Using another language, the c-liquid appears. One can interpret this second transition as follows. It is the exchange interaction that stabilizes the step-like density profile[39] because it provides the minimum of the exchange energy $H_{\rm ex}$ [Eq. (9)] for a given total magnetization. However, this kind of profile does not simultaneously minimize the Hartree (i.e., electrostatic) part of the energy $H_{\rm H}$ [Eq. (10)]. Electrostatics tends to create a smoothly varying density, and starting from the point where the width of the region "1" gets comparable with the range of the exchange interaction, which is $l_{\rm B}$ at the lowest Landau level, the exchange interaction is no longer able to counteract this tendency. With further softening of the confining potential (Fig. 5d), the c-liquid regions becomes much wider than the region "1"[41] (this regime was considered by Chklovskii et al[42]).

In our chess-board model the situation is similar: when the interaction strength is comparable or larger than that of the disorder, the narrow bulk edge channels along the sides of the chess-board we were talking about earlier in this Section, become wide compressible regions occupying a large portion of the total area. The saddle points undergo the same reconstruction. We are going to show below that for $N$ close to $N_c$ where $\delta\nu \simeq 1/2$, the size of the region "1" at the saddle points is much smaller than the size of the surrounding compressible areas (Fig. 6). In this case the profile of $\nu_N(\boldsymbol{r})$ in the cross-section going through the saddle point via labels "2"-"1"-"0" in Fig. 6a is similar to that shown in Fig. 5d.

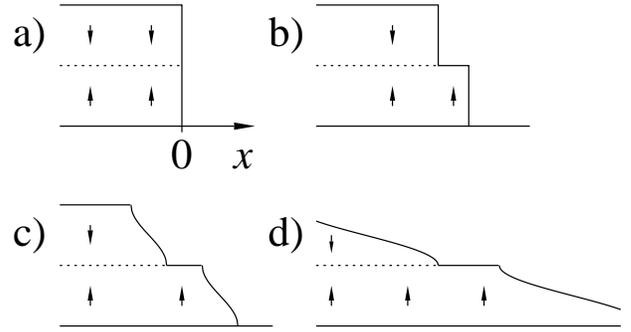

FIG. 5. The "0"-"2" to "0"-"1"-"2" transition in an edge of the wire[39]. (a) For a steep confining potential the edge is in "0"-"2" state. (b) As the confining potential is softened, the region "1" appears and grows. (c) When the width of the region "1" reaches $l_{\rm B}$, the c-liquid strips appear. (d) Eventually, the c-liquid strips become much wider than region "1".

The fact that c-liquid regions are wide has qualitatively new implications for the transport properties. The question now is whether or not this entire compressible area retains the metallic-type of conduction. Efros[15,21,43] suggested that yes, all the c-liquid is metallic. Since the percolation through the c-liquid is now achieved in wide ranges of average $\nu$ this suggestion lead him to the conclusion that the peaks in $\sigma_{\rm xx}$ must be wide, and correspondingly, the quantum Hall plateaux must be narrow[43] (see also Refs. 42, 44). Indeed, this is what is observed at relatively high temperatures[45]. However, it is well-known that at lower temperatures peaks become narrow. One interpretation of these phenomena was suggested in Ref. 42 and is as follows. As the temperature goes down, some fraction of the c-liquid with filling factors close to integer values becomes localized (or perhaps pinned); thus, the range of filling factors where c-liquid is metallic shrinks. If the want to preserve the property that only a single state per LLS is delocalized at zero temperature, then we have to assume that only very narrow strips of the c-liquid where $\nu(\boldsymbol{r})$ is half-integer remain metallic in the limit $T \to 0$. Therefore, we have to refine our criterion to have the conductivity peaks in the following way: the peaks corresponding to Landau level number $N$ appear when there is a percolation through $\nu_N(\boldsymbol{r}) = 1/2$ (for spin up peak) and $\nu_N(\boldsymbol{r}) = 3/2$ (for spin down peak). This is a very strong assumption, but we do not know an alternative way to interpret the low-temperature magnetoresistance data.

In our chess-board model the peaks in $\sigma_{\rm xx}$ corresponds to the configurations of $\nu_N = 1/2$ and $\nu_N = 3/2$ fillings at the saddle points (Fig. 6). The difference in the average filling factor for these configurations, which is $\delta\nu$, can be readily estimated as the ratio of the area occupied by the c-liquid near a single saddle point to the area of a chess-board cell, i.e., it is still given by Eq. (52) if by $\delta x$ we now understand the size of the *compressible* area. Then as in the large density case, $\delta\nu = 1/2$ corresponds to $\delta x \sim d \gg \xi$. Let us show that for such $\delta x$ the size of the incompressible region "1", which we denote by $a$ (see Fig. 6a), is much smaller than $\delta x$.

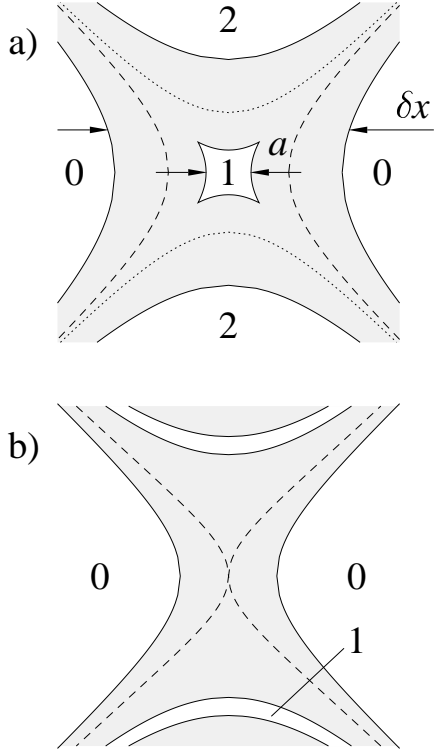

FIG. 6. The structure of a saddle point. Shown by grey is the area occupied by the c-liquid. Inside of this area two contours of constant filling factor, $\nu(\mathbf{r}) = 2N + 1/2$ (dashed line) and $\nu(\mathbf{r}) = 2N + 3/2$ (dotted line), are important for conductivity. (a) Average $\nu$ is $2N+1$ and the same is $\nu$ at the saddle point. Both contours are away from the saddle point. Thus, they do not percolate through the sample and $\sigma_{xx}$ is zero. (b) The average filling factor is decreased from the value of $2N+1$ by such an amount that $\nu$ at the saddle point becomes $2N + 1/2$. This brings the former contour to the percolation condition, while the latter contour is pushed further away from the saddle point. This corresponds to the spin up peak in $\sigma_{xx}$.

We can estimate $a$ in the way similar to the one used in Ref. 42 for the wire geometry. Without the exchange interaction the saddle point would be covered by a compressible region of size $\delta x$ and the external potential $V(\mathbf{r})$ would be completely screened so that the electrostatic energy be minimized. Due to the exchange interaction the region "1" appears. It can be considered as a quadrupole redistribution of the charge on top of the compressible region. This redistribution causes the deviation of the Hartree potential from its value for the complete screening case. The magnitude of the deviation is of the order $(e^2 n_{\mathrm{L}} a^3/\varepsilon_*)(\delta x)^{-2}$, where the last factor is the typical value of the second derivative of the filling factor at the saddle point (for the complete screening case). Now $a$ can be estimated from the condition that the deviation in the Hartree potential is of the same order of magnitude as the deviation in the exchange potential, which is $\alpha \hbar \omega_c$ if $a \gg \xi$. Using Eq. (44) for the effective dielectric constant $\varepsilon_*$ we find $a \sim (\delta x)^{2/3}(\alpha R)^{1/3}$. Replacing $\alpha R$ by $\xi$, we obtain

$$a \sim (\delta x)^{2/3}\xi^{1/3}. \quad (54)$$

Observe that for $\delta x \gg \xi$ we have $\xi \ll a \ll \delta x$, which makes our estimate consistent.

For $\delta\nu \simeq 1/2$ we have $\delta x \gg \xi$ and, therefore, $a \ll \delta x$ as we claimed above. In this case $\delta x$ as a function of $N$ can be found assuming that the entire area of the saddle point is occupied by the c-liquid as it would be without the exchange interaction. The evolution of the system at this stage is governed by the *Hartree* interaction (compare to Ref. 39), which makes the present case qualitatively different from the large density case where the exchange part of the interaction played the major rôle. The reason for this difference is that in the present case the exchange interaction is no longer enhanced through the collective exchange field of neighboring saddle points — they are too far apart.

Let us now calculate $\delta x$ as a function of $N$. Having done that, we will be able to find $N_c$ from the condition $\delta x \sim d$. Similar to the estimate of $a$ earlier in this Section, that of $\delta x$ will be also based on the fact that the charge distribution in the compressible region can be described as a quadrupole. In this case it is the quadrupole formed on top of the "0"-"2" state. Indeed, in the present state there are more electrons in the two "0" regions, so there is a negative charge there, and there are fewer electrons in the two "2" regions, so they hold some additional positive charge. The size of the quadrupole is such that the sum of the Hartree and the external potential energies has the lowest possible value. The energy of the quadrupole is, essentially, the product of its quadrupole moment and the second derivative of the external potential. Therefore, $\delta x$ can be estimated by equating the second derivatives of $V$ and the Hartree potential:

$$\frac{\partial^2 V}{\partial x^2} \sim \frac{e^2 n_{\mathrm{L}}}{\varepsilon_* \delta x}, \quad (55)$$

which gives

$$\delta x \sim \gamma d, \quad (56)$$

where $\gamma$ is the dimensionless parameter showing the relative strengths of the interaction and the external potential:

$$\gamma = \frac{1}{\partial^2 V/\partial x^2} \frac{e^2 n_{\mathrm{L}}}{\varepsilon_* d}. \quad (57)$$

Eqs. (52,56) show that $\delta\nu = 1/2$ is achieved when $\gamma$ is of order unity.

For the chess-board potential $\gamma$ is defined unambiguously because all the saddle points are identical. In the random system this is not the case, but it is reasonable to assume that the point $\delta\nu = 1/2$ also corresponds to $\gamma \sim 1$ if in the definition of $\gamma$ we use a *typical* value of $\partial^2 V/\partial x^2$, for which we take

$$\frac{\partial^2 V}{\partial x^2} = \frac{e^2 a_{\rm B} n_i^{1/2}}{d^3} \left(\frac{d}{R}\right)^{1/2} \quad (58)$$

[compare to Eq. (40)]. With the help of Eqs. (44,58) we can express $\gamma$ in terms of $n$, $n_i$, and $N$:

$$\gamma = \left(\frac{n}{n_i}\right)^{1/2} \left(\frac{k_{\rm F} d}{4\pi N}\right)^{3/2}, \quad (59)$$

and hence, $N_c$ is given by

$$N_c \sim k_{\rm F} d (n/n_i)^{1/3}, \quad (60)$$

which reproduces Eq. (4b) apart from the numerical coefficient. The coefficient was derived in the following way. We performed a computer simulation for the chessboard potential given by Eqs. (50,51) with the purpose to find the ground state of charged liquid interacting via a Coulomb law with the dielectric constant $\varepsilon_*$. We imposed an additional constraint that the density at every point is non-negative and does not exceed $2n_{\rm L}$. The results of the simulation depend only on two dimensionless parameters: $\gamma$ and $\nu = n/n_{\rm L}$. By varying the average density $n$ in the system we could vary the density at the saddle points in the resulting ground state, in particular, to make it equal to $n_{\rm L}/2$ and $3n_{\rm L}/2$. We then calculated $\delta\nu$ as the difference in $\nu$ corresponding to these values of $n$. We found that $\delta\nu = 1/2$ is achieved at $\gamma = 0.1$. Eventually, we substituted this number into Eq. (59) and obtained Eq. (4b). Note that the results for large and moderate densities [Eq. (4a) and Eq. (4b)] match at the crossover point defined by $n/n_i \sim \alpha^{-3}$, or more precisely, by Eq. (5).

Concluding this Section, we want to briefly discuss the case of small densities $n/n_i \ll 1$. This case is similar to the case of moderate densities with two exceptions. First, in this case $R \ll d$ and the averaged potential $V(\boldsymbol{r})$ coincides with the unaveraged one so that Eq. (40) has to be replaced by

$$\rho(0)^{-1} = \frac{\pi}{2} \frac{e^2 a_{\rm B} n_i^{1/2}}{d} \quad (61)$$

and Eq. (58) by

$$\frac{\partial^2 V}{\partial x^2} = \frac{e^2 a_{\rm B} n_i^{1/2}}{d^3}. \quad (62)$$

As a result, Eq. (56) gives

$$\delta x \sim \left(\frac{n}{n_i}\right)^{1/2} \frac{d^3}{R^2}. \quad (63)$$

The other difference from the moderate density case is that the amplitude of $V$ near the transition exceeds $\hbar\omega_c$ many times and $\nu(\boldsymbol{r})$ varies typically not within $2N$ and $2N+2$ but in a much wider interval. In this case the proper modification of Eq. (52) is

$$\delta\nu \sim \frac{\langle V^2 \rangle^{1/2}}{\hbar\omega_c} \left(\frac{\delta x}{d}\right)^2. \quad (64)$$

Together with Eq. (63) this leads to Eq. (4c). The case of small densities is interesting in the sense that small regions (with size of order $R$) near the saddle points occupied by the c-liquid determine, in fact, a large value of $\delta\nu$, which is a sample averaged quantity. So, $\delta\nu$ can be of order unity while the average magnetization is very small.

Our primary goal so far was to find $N_c$ where by definition $\delta\nu = 1/2$. Another interesting question is the behavior of $\delta\nu$ for $N > N_c$. For the moderate density case we can use Eqs. (52,56,59) to find that the transition has the tail in the form

$$\delta\nu(N) \sim (N_c/N)^3, \quad N > N_c. \quad (65)$$

Surprisingly, the same expression is valid for the small density case as well. In the derivation of Eq. (65) we assumed that $\delta x \gtrsim \xi$. Now the question is whether or not it holds for smaller $\delta x$ where the compressible region near the saddle point become more narrow than the region "1" that it surrounds. To answer this question we have to include the exchange interaction in our consideration. However, Eq. (65) will be modified by at most a numerical factor (absent in Eq. (65) anyway) because at distances $r \ll \xi$ the exchange interaction $G_{\rm ex}(r)$, like the Hartree interaction, is equivalent to the Coulomb law with the dielectric constant $\varepsilon_*$ [see Eqs. (15,44)].

Eventually, $\delta x$ reaches $l_{\rm B}$. At this point our approximation of the system of discreet electrons by continuous liquid breaks down. For the chess-board $\delta\nu$ should drop to zero. As for the random system system, there is a finite probability to find untypically soft saddle points (with small value of $\partial^2 V/\partial x^2$), which remain magnetized. Therefore, we can not exclude the possibility that $\delta\nu$ is non-zero but is very small.

Finally, we would like to remark that while for low-mobility samples or for high-mobility heterostructures with large electron densities it is possible to express $N_c$ in terms of the single-particle scattering time $\tau$, which is rather crude overall characteristic of the disorder, in the case studied in the present Section, $N_c$ is determined by the properties of some particular areas in the sample, namely, the saddle points of the disorder potential.

## V. CONCLUSION

In this paper we demonstrated that the spin-splitting of the quantum Hall conductivity peaks is determined by the competition between the disorder and the electron-electron interactions. We showed that in the mean-field approximation the distance $\delta\nu$ between the spin-resolved $\sigma_{xx}$ peaks measured in filling factor vanishes starting from a certain peak number $N_c$ if the bare Zeeman splitting is zero. A non-zero Zeeman splitting smears the transition but in GaAs devices this splitting is very small and, in principle, can be totally eliminated by applying pressure. We calculated the peak number $N_c$ where the spin-splitting disappears as a function of the heterostructure parameters and suggested a modified global phase diagram for the quantum Hall effect[25], which now includes spin (Fig. 2). Our calculations are in qualitative agreement with available experimental data[1,2,23]. However, to verify our predictions in detail, experiments on high-mobility gated samples where the electron density can be varied are desirable[24]. Another way to verify these predictions would be the experiments where the effective density of randomly situated donors can be varied independently of the electron density (see Ref. 17).

Our theory of the disorder-induced destruction of the many-body energy gap can be applied to other systems with additional degrees of freedom, examples of which can be multi-valley semiconductors or double quantum well structures. For instance, in Si MOSFET each branch of the phase diagram shown in Fig. 2 acquires an additional "valley" fork (see also an experimental phase diagram by Kravchenko et al[46]).

The critical behavior, i.e., the question how $\delta\nu$ goes to zero in the vicinity of the transition requires an additional investigation. Having in mind the experiments on gated samples where the electron density $n$ can be varied while keeping the peak number constant, we can define the proximity to the transition by the dimensionless parameter $(n - n_c)/n_c$, $n_c$ being the density at the very transition. Our mean-field theory, which applies to the case of large electron densities and not too small $\delta\nu$, gives

$$\delta\nu \propto \left(\frac{n - n_c}{n_c}\right)^{1/2}. \qquad (66)$$

For moderate and small densities the dependence is more complicated (see the discussion at the end of Sec. IV). Note that the experimental study of the critical behavior may be difficult in view of the finite temperature effects.

## ACKNOWLEDGMENTS

Useful discussions with I. L. Aleiner, L. I. Glazman, H.-W. Jiang, A. H. MacDonald, and D. G. Polyakov are greatly appreciated. We are grateful to I. L. Aleiner, L. I. Glazman, V. J. Goldman, H.-W. Jiang, E. Palm, L. P. Rokhinson, B. Su, and L. W. Wong for communicating to us their unpublished results. We thank A. A. Koulakov for help with the computer simulations.

This work was supported by NSF under Grant No. DMR-9321417.